\documentclass[%
 reprint,
 amsmath,amssymb,
 aip,
 superscriptaddress,
]{revtex4-2}

\usepackage{graphicx}
\usepackage{dcolumn}
\usepackage{physics}
\usepackage{booktabs}
\usepackage{xcolor}
\usepackage[colorlinks=true,linkcolor=black,anchorcolor=black,citecolor=black,filecolor=black,menucolor=black,runcolor=black,urlcolor=black]{hyperref}
\usepackage[colorlinks=true,linkcolor=black,anchorcolor=black,citecolor=black,filecolor=black,menucolor=black,runcolor=black,urlcolor=black]{hyperref}

\begin{document}

\preprint{APS/123-QED}

\title{Magnetoacoustic waves in a highly magnetostrictive FeGa thin film}

\author{Marc Rovirola}
\email{marc.rovirola@ub.edu}
\affiliation{Departament de Física de la Matèria Condensada, Facultat de Física, Universitat de Barcelona, 08028 Barcelona, Spain}
\affiliation{Institut de Nanociència i Nanotecnologia (IN2UB), Universitat de Barcelona, 08028 Barcelona, Spain}
\author{M. Waqas Khaliq}
\affiliation{Departament de Física de la Matèria Condensada, Facultat de Física, Universitat de Barcelona, 08028 Barcelona, Spain}
\affiliation{ALBA Synchrotron Light Source, 08290 Cerdanyola del Vallès, Spain}
\author{Blai Casals}
\affiliation{Institut de Nanociència i Nanotecnologia (IN2UB), Universitat de Barcelona, 08028 Barcelona, Spain}
\affiliation{Departament de Física Aplicada, Facultat de Física, Universitat de Barcelona, 08028 Barcelona, Spain}
\author{Adrian Begué}
\affiliation{Departamento de Física de Materiales, Facultad de Ciencias Físicas, Universidad Complutense de Madrid, 28040 Madrid, Spain}
\affiliation{Facultad de Ciencias, Universidad de Zaragoza, 50009 Zaragoza, Spain}
\author{Neven Biskup}
\affiliation{Departamento de Física de Materiales, Facultad de Ciencias Físicas, Universidad Complutense de Madrid, 28040 Madrid, Spain}
\author{Noelia Coton}
\affiliation{Departamento de Física de Materiales, Facultad de Ciencias Físicas, Universidad Complutense de Madrid, 28040 Madrid, Spain}
\affiliation{Instituto de Magnetismo Aplicado, UCM-ADIF-CSIC, Las Rozas 28232, Spain}
\author{Joan Manel Hernàndez}
\affiliation{Departament de Física de la Matèria Condensada, Facultat de Física, Universitat de Barcelona, 08028 Barcelona, Spain}
\affiliation{Institut de Nanociència i Nanotecnologia (IN2UB), Universitat de Barcelona, 08028 Barcelona, Spain}
\author{Miguel Angel Niño}
\affiliation{ALBA Synchrotron Light Source, 08290 Cerdanyola del Vallès, Spain}
\author{Michael Foerster}
\affiliation{ALBA Synchrotron Light Source, 08290 Cerdanyola del Vallès, Spain}
\author{Alberto Hernández-Mínguez}
\affiliation{Paul-Drude-Institut für Festkörperelektronik, Leibniz-Institut im Forschungsverbund Berlin e.V., 10117 Berlin, Germany}
\author{Rocío Ranchal}
\affiliation{Departamento de Física de Materiales, Facultad de Ciencias Físicas, Universidad Complutense de Madrid, 28040 Madrid, Spain}
\affiliation{Instituto de Magnetismo Aplicado, UCM-ADIF-CSIC, Las Rozas 28232, Spain}
\author{Marius V. Costache}
\affiliation{Departament de Física de la Matèria Condensada, Facultat de Física, Universitat de Barcelona, 08028 Barcelona, Spain}
\affiliation{Institut de Nanociència i Nanotecnologia (IN2UB), Universitat de Barcelona, 08028 Barcelona, Spain}
\author{Antoni García-Santiago}
\affiliation{Departament de Física de la Matèria Condensada, Facultat de Física, Universitat de Barcelona, 08028 Barcelona, Spain}
\affiliation{Institut de Nanociència i Nanotecnologia (IN2UB), Universitat de Barcelona, 08028 Barcelona, Spain}

\author{Ferran Macià}
\email{ferran.macia@ub.edu}
\affiliation{Departament de Física de la Matèria Condensada, Facultat de Física, Universitat de Barcelona, 08028 Barcelona, Spain}
\affiliation{Institut de Nanociència i Nanotecnologia (IN2UB), Universitat de Barcelona, 08028 Barcelona, Spain}

\date{\today}

\begin{abstract}
The interaction between surface acoustic waves and magnetization offers an efficient route for electrically controlling magnetic states. Here, we demonstrate the excitation of magnetoacoustic waves in galfenol, a highly magnetostrictive alloy made of iron (72\%) and gallium (28\%). We quantify the amplitude of the induced magnetization oscillations using magnetic imaging in an X-ray photoelectron microscope and estimate the dynamic magnetoelastic constants through micromagnetic simulations. Our findings demonstrate the potential of galfenol for magnonic applications and reveal that, despite strong magnetoelastic coupling, magnetic interactions and spin-wave dispersion relations significantly influence the overall amplitude of magnetoacoustic waves.

\end{abstract}

\maketitle

\section{Introduction}\label{sec:intro}

Integrating magnetic devices into electronic circuits requires efficient mechanisms for electrical control of magnetization in ferromagnetic (FM) nanomagnets, as opposed to magnetic field control. Electric-field control of magnetism has garnered significant attention for its potential to design ultralow-power spintronic devices.\cite{Matsukura2015} Current research focuses on using spin-polarized currents to exert torques on magnetic moments by transferring spin angular momentum.\cite{SLONCZEWSKI1996,Berger}  Various device geometries and material combinations have been proposed to generate high spin current densities, often leveraging spin-orbit interactions through Rashba, Dresselhaus, and spin Hall effects.\cite{PhysRev.100.580,Chernyshov2009, MihaiMiron2010,LiuScience,Wang2016} This approach allows for manipulating the magnetization in nanostructures with significantly reduced energy requirements.

Another efficient method for controlling magnetization using electric fields involves electrical tailoring magnetic anisotropy, a high-speed technique compatible with existing technologies.\cite{Chiba2008,Iurchuck_STRAIN} In this work, we study phonon coupling to magnetization \cite{Bukharaev2018straintronicRev,Foerster_2019} in bilayer heterostructures that combine a piezoelectric material with a highly magnetostrictive FM iron-gallium (Fe$_{72}$Ga$_{28}$) alloy. We use voltage-controlled surface acoustic waves (SAWs) to generate a spatially and temporally varying magnetic anisotropy in the FM alloy.

SAWs are MHz-GHz strain waves that propagate along the surface of a solid\cite{gangulyMagnetoelasticSurfaceWaves1976, fengMechanismInteractionSurface1982} and can originate from oscillating voltages through interdigital transducers (IDTs) in a piezoelectric material. These waves create an oscillating strain that can induce a varying magnetic anisotropy field in a magnetostrictive material.\cite{yang2021acoustic,puebla2022perspectives, SAW_roadmap_2019} The GHz frequency range is particularly relevant for magnetization dynamics because it matches internal resonance energies, and the associated spin waves have wavelengths in the nanometer-micrometer range.

\begin{figure*}[htb]
    \centering
    \includegraphics[width=1\textwidth]{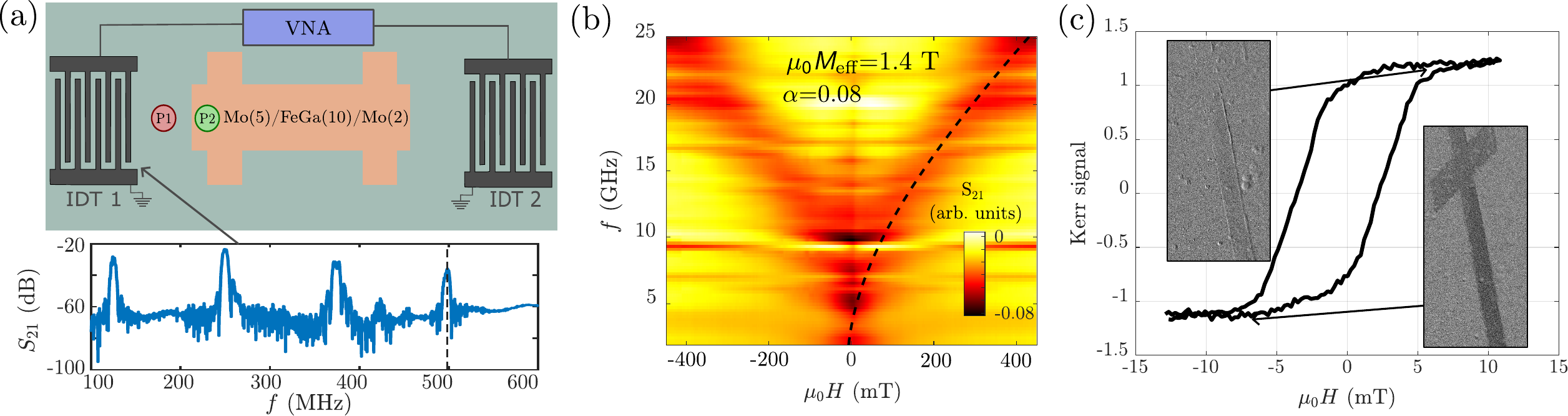}
    \caption{(a) The top panel shows a schematic illustration of the magnetoacoustic device setup, which includes a piezoelectric substrate (LiNbO$_3$) and a magnetostrictive thin film of FeGa (10 nm). This setup also features a Mo buffer layer (5 nm) and a Mo capping layer (2 nm). The bottom panel displays the transmission coefficient ($S_{21}$) of the interdigital transducers (IDTs), highlighting harmonics that start at 125 MHz. The dashed vertical line marks the frequency used in the experiments (500~MHz). (b) Ferromagnetic resonance data of the FeGa film used in the magnetoacoustic device. The vertical axis represents the microwave (MW) signal frequency applied to a coplanar waveguide, inducing uniform precession in FeGa. In contrast, the horizontal axis shows the externally applied magnetic field. The color scale indicates the relative loss of MW power after the RF field excites the FeGa film, with darker shades (black and red) indicating stronger absorption of MW power by the FeGa film. The black dashed line represents the fit of Kittel’s equation, from which we determine an effective magnetization saturation of $\mu_0M_{\mathrm{eff}} = 1.4$ T. (c) Magneto-optical Kerr effect (MOKE) hysteresis loop of the patterned FeGa, revealing a coercive field below 5 mT. The insets show MOKE images captured at opposite saturation points in the magnetic cycle.}
    \label{setup}
\end{figure*}

In the past decade, numerous studies have demonstrated that SAWs can efficiently couple to magnetization via the magnetoelastic effect,\cite{JMH_SAW_APL_2006,weiler2011elastically, dreherSurfaceAcousticWave2012, Weiler2012spin, gowthamTravelingSurfaceSpinwave2015b, Labanowski2016,seemann2022magnetoelastic} focusing on the magnetoacoustic waves (MAWs) generated by SAWs.\cite{Kuszewski_2018, Kuszewski_2018, McCord_AEM_2022,casalsGenerationImagingMagnetoacoustic2020,rovirola2023_physrevapp,khaliq2023antiferromagnetic,rovirola2023study} Transition metal FMs, including Heusler alloys with similar bulk magnetostriction coefficients ($\sim 10^{-5}$), have shown significant variations in MAW amplitudes. 

Previous studies in galfenol with lower Ga content ($18-20 \%$) have used SAWs to reduce the coercive field in thin films\cite{Li_ieee_FeGa_SAW} and create magnetic patterns \cite{Li_JAP_FeGa_SAW} at the microscale, demonstrating the potential for controlling magnetic states with oscillating strain. Field sensors have been proposed based on the field-dependent SAW absorption observed in thick FeGa films.\cite{Li_SAW_FeGa, Sun_SAW_FeGa} However, direct measurements of magnetoelastic waves and the quantification of magnetoelastic coefficients in the GHz regime are still lacking.

In this work, we investigate the interaction of SAWs with magnetization in a 10-nm-thick Fe$_{72}$Ga$_{28}$ thin film, 
which has rich-Ga content compared with previous studies,\cite{Ruffoni_prl2008, Gopman2017_FeGa, Duquesne_2018, Li_ieee_FeGa_SAW, Li_JAP_FeGa_SAW, Li_SAW_FeGa, Sun_SAW_FeGa} and falls within the `second peak' of magnetostriction for FeGa alloys, a region known for enhanced magnetostrictive properties.\cite{XING20084536_Ga_content, Bartolome2020}
We chose the thickness of 10 nm to ensure a low coercive field and small anisotropy.\cite{Gopman2017_FeGa}

Our experiment involves placing the FM thin film on the acoustic path of a piezoelectric substrate and exciting SAWs through IDTs at 500 MHz. Due to the magnetostrictive nature of FeGa, the deformation caused by the SAWs generates an oscillating anisotropy, leading to periodic changes in the magnetization direction. We directly image the resulting MAWs using X-ray microscopy and quantify the dynamic magnetoelastic coupling.

\section{Experimental details}

The sample under study consists of a piezoelectric 127.86 deg Y-cut LiNbO$_3$ substrate and a 10-nm-thick FeGa film deposited by DC magnetron sputtering and patterned like a Hall bar [see Fig. 1(a)]. We deposited 5-nm thick buffer and 2-nm thick capping molybdenum layers below and above the galfenol film to ensure good adhesion and prevent oxidation. The sample's stoichiometry is Fe$_{72}$Ga$_{28}$, but it will be referred to as FeGa throughout this article. More details on FeGa growth can be found elsewhere.\cite{Bartolome2020}
Unidirectional IDTs were patterned at opposite ends of the FeGa film using electron beam lithography, metal evaporation, and liftoff [see Fig \ref{setup}(a)]. This configuration allows both the generation of SAWs that propagate along the FM film and the measurement of transmitted SAW power ($S_{21}$ coefficient). We designed the IDTs to generate Rayleigh SAWs, characterized by in-plane and out-of-plane longitudinal strain components ($\varepsilon_{xx}$ and $\varepsilon_{zz}$, respectively) and a shear strain component ($\varepsilon_{xz}$). The spatial periodicity of the finger-like electrodes in the IDTs was designed to excite SAWs with a fundamental frequency of 125 MHz and its corresponding harmonics [see the $S_{21}$ spectrum in Fig. \ref{setup}(a)]. Through the magnetoelastic effect, the strain waves from the SAWs are transformed into a dynamically effective magnetic field, producing the torque on the magnetization of FeGa and thus generating MAWs.

\section{Results}
We initially characterized the sample using ferromagnetic resonance (FMR) and magneto-optical Kerr effect (MOKE) microscopy. We measured the FMR on a separate FeGa sample grown simultaneously with the magnetoacoustic device on an identical substrate to ensure that both objects had the same magnetic properties. In the FMR experiment, we placed the sample on a coplanar waveguide and excited it at various frequencies while sweeping the magnetic field from $-400$ mT to $400$ mT. At specific magnetic fields and MW frequencies, the magnetization enters resonance, absorbing more MW power and resulting in a dip in the transmission coefficient, which appears as darker shades in Fig. \ref{setup}(b). The frequency and magnetic field of these resonances follow Kittel's equation\cite{Kittel} 
\begin{equation}
    f=\gamma\sqrt{\mu_0H(\mu_0H+\mu_0M_{\text{eff}})} \, ,
    \label{eq:kittel}
\end{equation} 
where $\gamma$ is the gyromagnetic ratio ($\gamma\approx 28$\ GHz/T), and $\mu_0M_{\mathrm{eff}}$ is the effective magnetization saturation, which we estimated to be around 1.4~T from data shown in Fig.\ \ref{setup}(b). The effective magnetization saturation is given as $M_{\mathrm{eff}} = M_{S}-H_{k}$, where $M_{S}$ is the magnetization saturation, and $H_{k}$ is the out-of-plane anisotropy field. We expect this field to be small in FeGa, so that $M_{\rm eff} \approx M_{S}$. In addition, the linewidth ($\mu_0 \Delta H$) of the FMR resonances follows a linear frequency dependence \cite{Kalarickal}
\begin{equation}
    \mu_0 \Delta H = \frac{2}{\gamma}\alpha f + \mu_0 \Delta H_{\rm{in}},
\end{equation}
\noindent where $\alpha$ is the Gilbert damping parameter, indicating the rate of energy dissipation, and $\mu_0 \Delta H_{\rm{in}}$ represents the inhomogeneous linewidth broadening, which accounts for magnetic inhomogeneities in the sample. Our FeGa thin films have broad FMR peaks, resulting in a high damping value of about $\alpha = 0.08$.

\begin{figure}[ht]
    \centering
    \includegraphics[width=01\columnwidth]{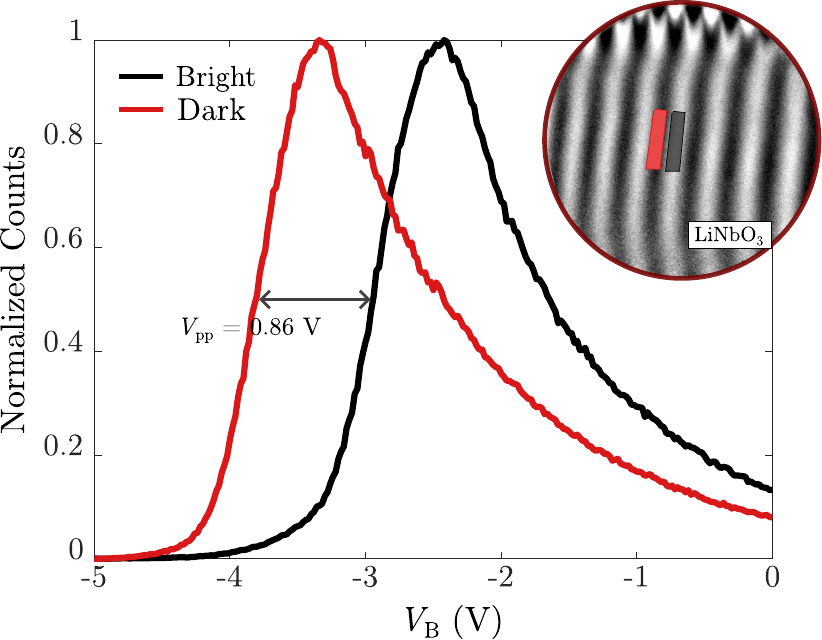}
    \caption{Scans of bias voltage, $V_{\rm B}$, for opposite SAW phases. The red line represents the dark SAW phase, while the black line represents the bright SAW phase. The horizontal displacement between the spectra is the peak-to-peak voltage of the SAW at the sample surface, $V_{\rm pp}$. The inset shows an XPEEM image at location P1 on LiNbO$_3$. The black and red rectangles are the regions where we extract the curve data as we sweep the bias voltage.}
    \label{STVscanSAW}
\end{figure}

\begin{figure*}[ht]
    \centering
    \includegraphics[width=0.97\textwidth]{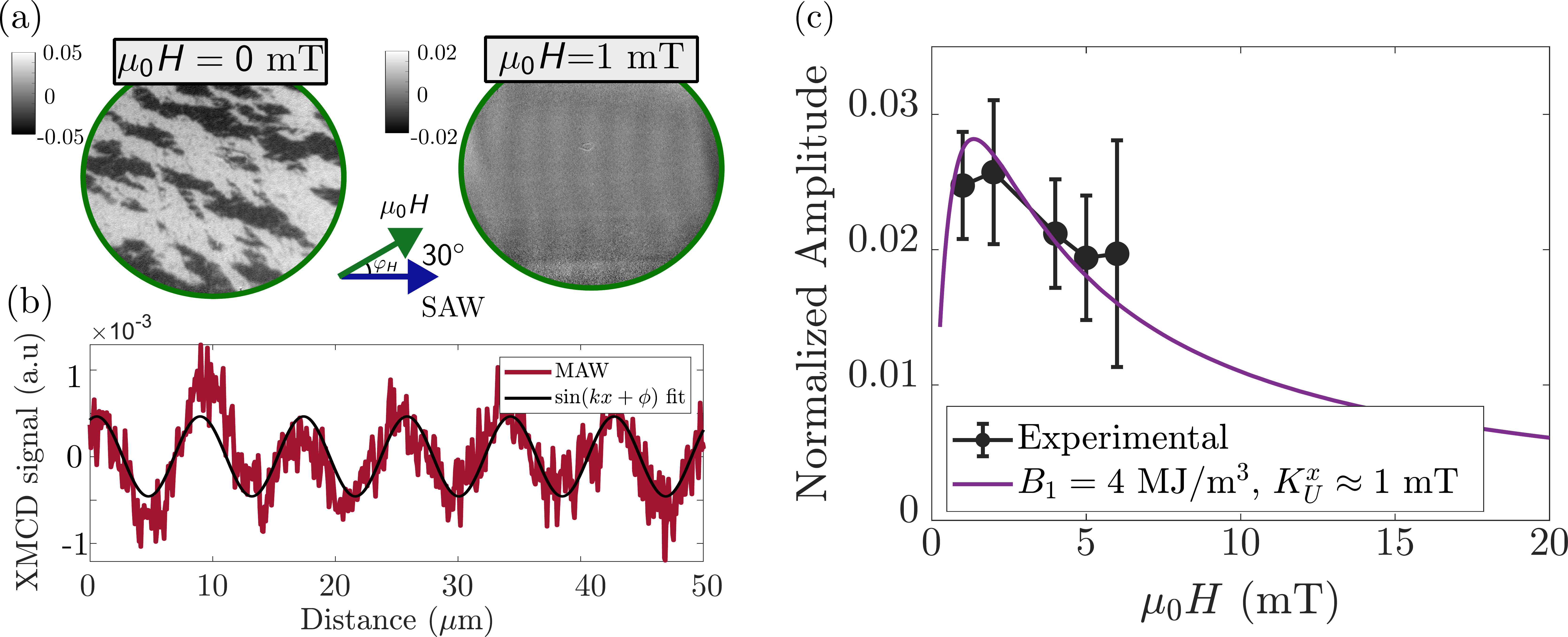}
    \caption{(a) The left panel displays an XMCD image showing magnetic domains on FeGa. The right panel displays a 2-phase XMCD image highlighting MAWs on FeGa at a magnetic field of 1 mT. (b) Spatial MAWs profile (red curve) at a fixed magnetic field of 1 mT. The black curve is the fitted sinusoidal function. (c) Magnetic field dependence of the normalized MAW amplitude derived from the XMCD profiles (connected black dots; see the text for details). The purple continuous line represents the micromagnetic simulation results that better reproduce the experimental results, with $B_1=4$ MJ/m$^3$, with an in-plane anisotropy in the $x$-direction (parallel to the SAW propagation direction) of 1 mT.
    }
    \label{MAW_XMCD}
\end{figure*}

Figure \ref{setup}(c) shows the in-plane magnetic loop of FeGa obtained through MOKE microscopy. The loop exhibits magnetic hysteresis with a coercive field of less than 5 mT and full magnetic saturation at around 7 mT. The insets display MOKE images of the Hall bar under saturated conditions, although imaging the small magnetic domains with MOKE proved challenging.

To study the effect of SAWs on the magnetization dynamics of FeGa, we used direct imaging with X-ray photoemission electron microscopy (XPEEM) at the ALBA synchrotron facility. This technique allows us to observe and quantify both SAWs and MAWs. The experiment involves illuminating the sample with X-rays at the $L_3$ energy edge of Fe, which excites secondary electrons. A 10 kV potential between the device and the XPEEM objective pulls out these electrons. Two types of images can be obtained: XPEEM, which is sensitive to the piezoelectric voltage at the sample surface, and X-ray magnetic circular dichroism (XMCD), which provides magnetic contrast by subtracting two images taken with opposite circular helicity, canceling non-helicity-dependent effects.

SAWs are emitted by the IDTs and propagate along the acoustic path. The AC electric field applied to the IDT to generate SAWs is synchronized with the repetition rate of the synchrotron X-ray bunches, enabling stroboscopic imaging at specific phases of the SAW.\cite{foerster2017direct} The local piezoelectric field from the SAW affects the kinetic energy of the photoelectrons that are accelerated out from the piezoelectric sample and is highly sensitive to the electric field between the sample and the microscope objective. By adding a bias voltage ($V_\mathrm{B}$) to the sample relative to the objective, we can change the contrast of the SAWs. Scanning the detector counts as a function of $V_\mathrm{B}$ at opposite SAW phases allows us to measure the peak-to-peak amplitude of the piezoelectric potential at the sample surface.

The inset of Fig. \ref{STVscanSAW} shows an XPEEM image of the LiNbO$_3$ substrate at location P1 [see Fig. \ref{setup}(a)]. The red and black rectangles represent the studied areas, corresponding to opposite SAW phases. The spectra in Fig. \ref{STVscanSAW} display normalized detector counts as a function of $V_\mathrm{B}$ for each phase, with the colors matching the studied areas in the inset. Subtracting both spectra, we obtain the SAW peak-to-peak voltage ($V_\mathrm{pp}$), the horizontal shift between both spectra. We can determine the corresponding strain amplitudes by solving the coupled electromagnetic and elastic equations.\cite{1973AcousticFA_Auld} In our experiment, we measured $V_\mathrm{pp} = 0.86$ V at 500~MHz and calculated the strain component amplitudes as $\varepsilon_{xx} = 6.4 \times 10^{-5}$, $\varepsilon_{zz}=1.0 \times 10^{-5}$, and $\varepsilon_{xz} = 1.4 \times 10^{-5}$.

To study the MAWs excited in the FeGa film, we took a series of XMCD images at position P2 [see Fig. \ref{setup}(a)] under various magnetic field strengths. To eliminate static features such as magnetic domains and surface impurities at low magnetic fields, we used a four-image process: taking two images with the same X-ray helicity at a $180^\circ$ phase difference in SAW and subtracting them, then obtaining magnetic contrast by subtracting images taken with opposite helicity. We call this a two-phase XMCD image. Figure \ref{MAW_XMCD}(a) shows a standard XMCD image of magnetic domains (left) and a two-phase XMCD image of MAWs (right). From the MAW images, we extracted their profile and fitted it to a sinusoidal wave, using amplitude and phase as fitting parameters [see Fig. \ref{MAW_XMCD}(b)]. To quantify the amplitude, we normalized it by the contrast difference between opposing domains, which is proportional to the magnetization saturation ($M_S$). This normalized amplitude is shown in Fig. \ref{MAW_XMCD}(c), with each black point indicating the amplitude at different magnetic fields, ranging from 1 to 6 mT. The magnetic field is applied in the film plane at a $30^\circ$ angle to the SAW propagation direction to maximize the magnetoelastic torque.\cite{Puebla_2020, weiler2011elastically, dreherSurfaceAcousticWave2012} Our MAW quantification estimates a magnetization precession angle of $1^\circ$ to $1.5^\circ$, with a maximum efficiency of $2.3^\circ$/strain, comparable to nickel and cobalt.\cite{khaliq2023antiferromagnetic}

Next, we compared our results to micromagnetic simulations to quantify the magnetoelastic coupling constant. We used the MuMax3 simulation software,\cite{mumax3} which incorporates all effective field contributions, including Zeeman, demagnetization, anisotropy, and magnetoelastic fields generated by the coupling between SAWs and magnetization. The effective magnetoelastic field is given by
\begin{align}\label{eq:Fme}
   \nonumber F_{\text{me}}= &\,\, B_1\left[\varepsilon_{x x} m_x^2+\varepsilon_{y y} m_y^2+\varepsilon_{z z} m_z^2\right] \\
    &+2 B_2\left[\varepsilon_{x y} m_x m_y+\varepsilon_{x z} m_x m_z+\varepsilon_{y z} m_y m_z\right],
\end{align}

\noindent where $B_{i}$ are the magnetoelastic constants, and $m_{i}$ are the normalized magnetization components. The sample is polycrystalline, so $B_1 = B_2$. The simulation that better reproduces the experimental results uses $B_1 = 4$~MJ/m$^3$ and a small in-plane anisotropy of about 1 mT in the $x$ direction, parallel to the direction of SAW propagation. The results, shown as a solid purple line in Fig. \ref{MAW_XMCD}(c), reveal a peak at 1 mT due to the in-plane uniaxial anisotropy rather than a magnetic resonance. Since the SAW frequency is below the resonance frequency, the maximum effective field is near zero and thus generates maximum magnetization oscillation.


We chose FeGa films for their high magnetostriction, anticipating large MAW amplitudes. However, from the experimental results we estimate a value of the magnetoelastic constant that is similar to those observed in nickel and cobalt thin films \cite{rovirola2023study} but with a MAW amplitude about ten times smaller than in nickel, though comparable to cobalt. We ascribe this reduction in MAW amplitude to the $M_S$ value of FeGa and the dependence of MAW dispersion relations on the applied field angle. It is worth noticing that MAWs propagate at a certain angle relative to the magnetization direction (30$^{\circ}$ in the present study).

\begin{figure}[ht]
    \centering
    \includegraphics[width=0.9\columnwidth]{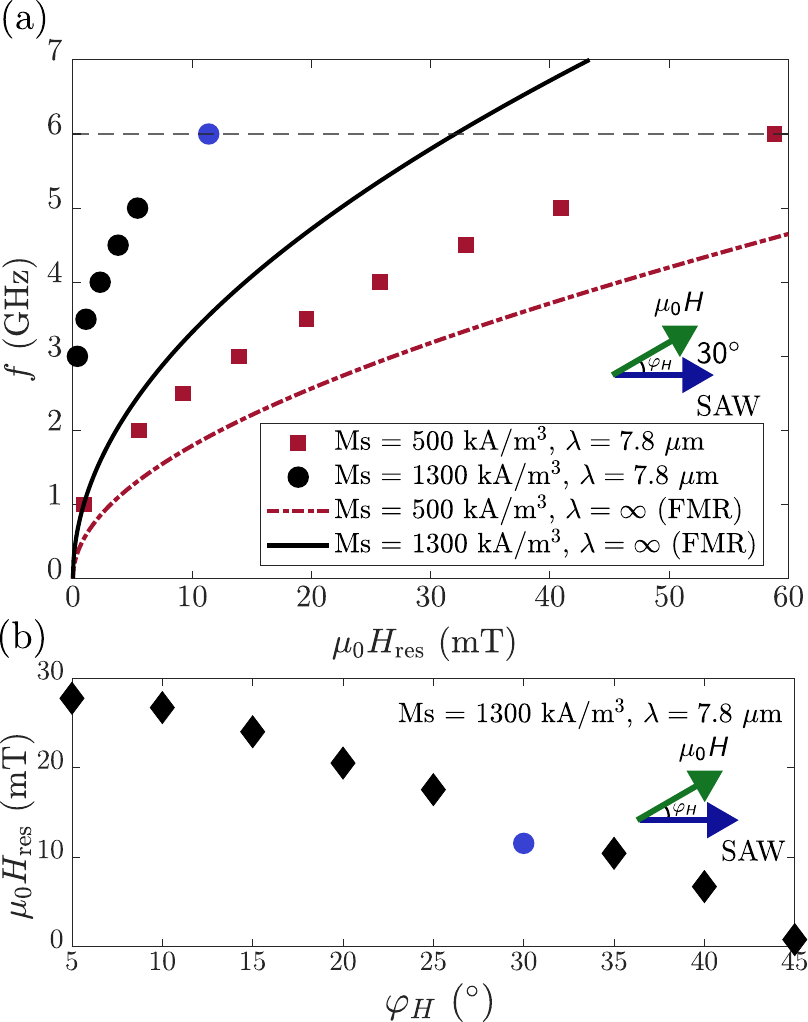}
    \caption{(a) Simulation results from MuMax3 showing the dependence of frequency on resonance magnetic field for two samples with different $M_s$ values. Solid and dashed-dotted lines represent FMR resonances for $M_s = 1300$ kA/m$^3$ and $M_s = 500$ kA/m$^3$, respectively. The dots and squares correspond to the MAW resonance condition given by an applied magnetic field and a SAW frequency when the SAW propagation forms an angle of $30^\circ$ with the applied field for $M_s = 1300$ kA/m$^3$ and $M_s = 500$ kA/m$^3$, respectively. (b) Simulation results from MuMax3 displaying the angular dependence of the resonant field of FeGa ($M_s = 1300$ kA/m$^3$) for an excitation frequency of 6 GHz when the angle between the SAW and the applied field is decreased from 45$^\circ$ to 0$^\circ$ in 5$^\circ$ intervals.}
    \label{simulations}
\end{figure}

According to Eq.\ (\ref{eq:kittel}), a high $M_S$ value at a fixed magnetic field shifts the resonance to higher frequencies. Figure \ref{simulations}(a) illustrates this by comparing the solid black curve (representing FeGa with $M_S = 1300$ kA/m$^3$) with the dashed-dotted red curve (representing nickel with $M_S = 500$ kA/m$^3$). The SAWs introduce a wavelength parameter in the magnetic waves, $\lambda=f_{\text{SAW}}/v_{\text{s}}$, where $f_{\text{SAW}}$ is the excitation frequency and $v_{\text{s}}$ is the speed of sound in the substrate. The non-uniform magnetization precession affects exchange and dipolar energies, typically leading to an additional increase in frequency with decreasing wavelength.\cite{slavin_dispersion}
Symbols in figure \ref{simulations}(a) show resonance frequencies for MAW with its corresponding wavelength ($\lambda=7.8$ $\mu$m) when the SAW propagation direction and the applied magnetic field form a $30^\circ$ angle ($\varphi_H$). Each point represents the MAW resonance frequency for a given magnetic field. For $M_S=500$ kA/m$^3$ (red squares), the resonance starts close to 500 MHz and increases with the applied magnetic field. In contrast, for $M_S = 1300$ kA/m$^3$ (black dots), the resonance starts around 3 GHz, so exciting it with SAWs at 500 MHz results in non-resonance MAWs. The gap between FMR ($\lambda = \infty$) and MAW resonance increases with $M_S$, as shown for nickel ($M_S=500$ kA/m$^3$) and FeGa ($M_S = 1300$ kA/m$^3$). 

The angle between the SAW propagation and the applied magnetic field, $\varphi_H$, also affects the size of the gap, increasing it as $\varphi_H$ grows. Figure \ref{simulations}(b) presents additional simulations for FeGa, showing the MAW resonant magnetic field ($\mu_0 H_{\rm res}$) for a 6 GHz SAW excitation [blue dot in panel (a)] as a function of $\varphi_H$. The resonance field decreases from nearly 30 mT when the SAWs and the applied field are parallel to almost zero at $\varphi_H = 45^\circ$.

\section{Conclusions}\label{section:conclusions}

To summarize, we investigated the dynamic interaction between SAWs and magnetization in a galfenol thin film on a piezoelectric LiNbO$_3$ substrate. We demonstrated the electric control of magnetization waves through XMPEEM measurements. Our findings revealed the presence of both SAWs and magnetoacoustic waves (MAWs) in the FM thin film, with the MAW amplitude being comparable to that in cobalt but smaller than in nickel. Our technique allows quantification of strain and MAW amplitude and thus enables us to obtain an estimate for the magnetoelastic constant ($B_1=4$~MJ/m$^3$) in sputtered 10 nm galfenol thin film with rich-Ga content (Fe$_{72}$Ga$_{28}$). Our simulations indicate that we generate MAW in a non-resonance mode and suggest that we can achieve larger MAW amplitudes when resonance conditions are fulfilled. The simulations also show that parameters such as the film's magnetic moment or propagation angle between MAW and magnetization play a crucial role in determining the resonance condition and setting the overall MAW amplitude.

\begin{acknowledgments}
The experiments were performed at CIRCE beamline at ALBA Synchrotron with the collaboration of ALBA staff. MR, MWK, BC, MVC, AGS, JMH, and FM acknowledge funding from MCIN/AEI/ 10.13039/501100011033 through grant number PID2020-113024GB-100 and EU NextGenerationEU/PRTR through Grant No. PDC-2023-145910-I00. MR is supported by FPI Grant PRE2021 -097235. AB, NC and RR thank project PID2021-122980OB-C51 (AEI/FEDER) of the Spanish Ministry of Science and Innovation. AB would like to acknowledge the funding received from the Ministry of Universities and the European Union-Next Generation for the Margarita Salas fellowship.
\end{acknowledgments}

\bibliographystyle{aip}

\end{document}